\documentclass[10pt]{article}

\usepackage{amsmath,amssymb,dsfont,slashed,color,hyperref}
\usepackage{authblk}

\usepackage[toc,page]{appendix}
\usepackage{graphicx}
\usepackage[shortlabels]{enumitem}

\definecolor{armygreen}{rgb}{0.29, 0.33, 0.13}

\newcommand{\half}{\frac{1}{2}}

\newcommand{\JJ}{\mathds{J}}
\newcommand{\KK}{\mathds{K}}
\newcommand{\DD}{\mathds{D}}
\newcommand{\TT}{\mathds{T}}
\newcommand{\ZZ}{\mathds{Z}}

\newcommand{\QQ}{\mathds{Q}}
\newcommand{\QQb}{\overline{\mathds{Q}}}

\newcommand{\GG}{\mathds{G}}

\newcommand{\AAA}{\mathds{A}}
\newcommand{\FF}{\mathds{F}}
\newcommand{\calF}{\mathcal{F}}
\newcommand{\calG}{\mathcal{G}}
\newcommand{\calH}{\mathcal{H}}
\newcommand{\calL}{\mathcal{L}}
\newcommand{\calR}{\mathcal{R}}
\newcommand{\calX}{\mathcal{X}}
\newcommand{\calXbar}{\overline{\mathcal{X}}}
\newcommand{\calY}{\mathcal{Y}}
\newcommand{\calYbar}{\overline{\mathcal{Y}}}
\newcommand{\se}{\slashed{e}}
\newcommand{\sT}{\slashed{T}}

\newcommand{\lD}{\overleftarrow{D}}

\newcommand{\psibar}{{\overline{\psi}}}
\newcommand{\chibar}{{\overline{\chi}}}
\newcommand{\tgamma}{\tilde{\gamma}}
\newcommand{\gf}{\gamma_5}

\newcommand{\wtX}{{\widetilde{X}}}
\newcommand{\calW}{{\mathcal{W}}}

\newcommand{\uantof}{Departamento de Física, Universidad de Antofagasta, Aptdo. 02800, Chile}

\newcommand{\cecs}{Centro de Estudios Cient\'{\i}ficos (CECs), Arturo Prat 514, Valdivia, Chile\\
Universidad San Sebasti\'an, General Lagos 1163, Valdivia, Chile
}

\begin{document}

\title{Embedding of the Georgi-Glashow $SU(5)$ model in the superconformal algebra}
\author[1]{P. D. Alvarez \thanks{E-mail: \href{mailto:pedro.alvarez@uantof.cl}{\nolinkurl{pedro.alvarez@uantof.cl}}}}

\author[1]{R. A. Chavez \thanks{E-mail: \href{mailto:rafael.chavez.linares@uantof.cl}{\nolinkurl{rafael.chavez.linares@uantof.cl}}}}
\affil[1]{\uantof}

\author[2]{J. Zanelli \thanks{E-mail: \href{mailto:z@cecs.cl}{\nolinkurl{z@cecs.cl}}}}
\affil[2]{\cecs}

\maketitle

\begin{abstract}
 We present a scheme to construct grand unified models based on the superconformal algebra and the inclusion of matter fields in the adjoint representation of supersymmetry. As an illustration, we implemented the Georgi-Glashow $SU(5)$ model. The model predics the existence of a hidden $(\mathbf{1},\mathbf{24},0)  + (\mathbf{5},\mathbf{5}^\ast,-y') + (\mathbf{5}^\ast,\mathbf{5},y')$ sector and an anomalous $U(1)_Z$.
\end{abstract}

\tableofcontents

%%%%%%%%%%%%%%%%%%%%%%%
\section{Introduction}\label{intro}

The simplest grand unified theory (GUT) is the Georgy-Glashow (GG) $SU(5)$ model, where the fifteen leptons and quarks of the Standard Model are included in the $\mathbf{5}^\ast$ and a $\mathbf{10}$ representation of $SU(5)$ \cite{Georgi:1974sy,Langacker:1980js} (in the one-family approximation). Among the welcomed features we have unification of coupling constants, quark charge quantization and a rationale for anomaly cancellation. The price to pay for such beauty is the presence of proton decay channels and the doublet-triplet splitting problems on top of the usual quadratic mass renormalization of the scalars.

There are improved models that address some of the issues of the GG model. To protect the Higgs doublets from quadratic mass renormalization, an $SU(5)$ model with softly broken supersymmetry was proposed \cite{Dimopoulos:1981zb}. To avoid proton decay and the doublet-triplet splitting problems of the GG $SU(5)$ model, the flipped $SU(5)$ model was proposed \cite{Barr:1981qv}. The supersymmetric flipped $SU(5)$ model, that produces hierarchical neutrino masses, was proposed \cite{Antoniadis:1987dx,Ellis:1992zr,Ellis:1993ks,Ellis:1992nq}. Flipped $SU(5)$ can also be embedded in string theory \cite{Antoniadis:1988tt,Campbell:1987gz,Lopez:1992kg}.

Minimal flipped $SU(5)$ GUT was shown to survive experimental electroweak limits \cite{Ellis:2002vk}. The flipped $SU(5)$ GUT has also been embedded in no-scale supergravity models \cite{Ellis:2017jcp}. Such model can generate no-scale inflation, similar to Starobinsky's inflation, \cite{Ferrara:2013rsa}, and makes predictions for cosmic microwave background observables that can be used to constrain significantly the parameters of the model.

One shortcoming of the flipped $SU(5)$ model is that the hypercharge is in the factor of the gauge group proposed for unification $(SU(5) \times U(1)_\chi)/\mathbf{Z}_5$. The fact that the structure $SU(n) \times U(1)$ can be found in the superconformal algebra is suggestive. However, the embedding of supersymmetric $SU(n)$ GUT models in the superconformal algebra faces technical difficulties \cite{Ferrara:1977ij,Kaku:1977rk}.

In this paper we present a model that circumvents those difficulties by using a novel implementation of supersymmetry with matter in the adjoint representation of the superconformal algebra. We use two basic assumptions, firstly we use a diagonal symmetry group 
\begin{equation}\label{diagsymm}
 SU(2,2|n)_\text{diag} = [SU(2,2|n) \times SU(2,2|d_n)]_\text{diag}
\end{equation}
where $d_n=n(n-1)/2$. Such a value comes from the embedding of supercharges in the superconformal algbera carrying a bi-fundamental index \cite{Alvarez:2021lda}. Secondly, we define a dual operator which is naturally embedded in $SU(2,2|n)$ \cite{Alvarez:2021zsw}. This dual operator induces a grading in the conformal algebra. This grading allows defining actions with gauge symmetry $G^+ \subset SU(2,2)\times SU(n) \times U(1)$, where $G^+$ is the group generated by the grading-even generators \cite{Alvarez:2021zsw}. The symmetries generated by the grading-odd generators, and the supersymmetries become on-shell symmetries \cite{Alvarez:2021qbu}. For the sake of simplicity we focus on the implementation of the GG model.

\section{The model}\label{model}

Let us define a model for gauge potential for the group (\ref{diagsymm}). The gauge connection is
\begin{align}
 \AAA =& \Omega+\QQb^i \se \psi_i + \psibar^i \se \QQ_i\,,\\
 \AAA' =& \Omega'+\half \QQb^{ij} \se \chi_{ij} + \half \chibar^{ij} \se \QQ_{ij} \,,
\end{align}
where
\begin{align}
\Omega =& \half \omega^{ab}\JJ_{ab}+f^a \JJ_a+g^a\KK_a+h\DD+A^I \TT_I+A\ZZ \,,\\
\Omega' =& \half \omega'^{ab}\JJ_{ab}+f'^a \JJ_a+g'^a\KK_a+h'\DD+A'^X \TT_X+A'\ZZ \,,
\end{align}
and $a=0,\cdots,3$, $I=1,\cdots, n^2-1$, $X=1,\cdots, d_n^2-1$. We are interested in the embedding of the Georgi-Glashow model into the superconformal algbera. For this purpose we will consider (\ref{diagsymm}) with $n=5$. Note that, in the present model, the gauge bosons $A$, $A'$ and $A'^{\widetilde{X}}$ with $\widetilde{X} = n^2, n^2+1, \cdots, d_n^2-1$, are new with respect to the Georgi-Glashow model. The spinors transform in the appropriate representations of $SU(5)$,
\begin{equation}
 (\psi^i)_L = \mathbf{5}^*\,, \quad (\chi_{ij})_L = \mathbf{10}\,. \label{repGG}
\end{equation}
with $\chi_{ij} = - \chi_{ji}$, and the fifteen left-handed quarks and lepton of the standard model are placed in a reducible $\mathbf{5}^\ast + \mathbf{10}$ representation,
\begin{align}
 \mathbf{5}^\ast =& (\mathbf{3}^\ast,\mathbf{1},\tfrac{1}{3})+(\mathbf{1},\mathbf{2}^\ast,-\tfrac{1}{2})\,,\\
 \mathbf{10} =& (\mathbf{3}^\ast,\mathbf{1},-\tfrac{2}{3})+(\mathbf{3},\mathbf{2},\tfrac{1}{6})+(\mathbf{1},\mathbf{1},1)\,,
\end{align}
where $(\mathbf{n}_3,\mathbf{n}_2,y)$ give the representations under $SU(3)$, $SU(2)$ and the weak hypercharge \cite{Langacker:1980js}.

The field assignment is as follows
\begin{equation}\label{ferr5s}
 (\psi^i)_L=\left(\begin{array}{c}
                     d^{c 1} \\
                     d^{c 2}  \\
                     d^{c 3}  \\
                     e^- \\
                     -\nu_e \\
                   \end{array}\right)_L\,,
\end{equation}
and
\begin{align}\label{ferr10}
  (\chi_{ij})_L= &\frac{1}{\sqrt{2}} \left(\begin{array}{ccccc}
                           0 & u^{c 3} & -u^{c 2} & -u_1 & -d_1 \\
                           -u^{c 3} & 0 & u^{c 1} & -u_2 & -d_2 \\
                           u^{c 2} & -u^{c 1} & 0 & -u_3 & -d_3 \\
                           u^1 & u^2 & u^3  & 0 & -e^+ \\
                           d^1  & d^2  & d^3 & e^+ & 0 \\
                         \end{array}\right)_{L}\,.
\end{align}

The action principle is based on a relative of the MacDowell-Mansouri action \cite{MacDowell:1977jt},
\begin{equation}\label{action}
 \mathcal{S}= - \int \left( \langle \xi \FF \circledast \FF \rangle + \langle \xi' \FF' \circledast \FF' \rangle \right)\,.
\end{equation}
The field strength is defined as usual by $\FF = d\AAA +\AAA \wedge \AAA$, where wedge products are assumed. The explicit expressions for the curvature components are in appendix \ref{curvaturesApp}. In the action (\ref{action}) we use a generalized dual operator defined by\footnote{It would be interesting to see how this Hodge operator relates to the one defined on a supermanifold \cite{CCG1,CCG2}.}
\begin{align}
\circledast \FF=&(\varepsilon_s S)\left(\half \calF^{ab} \JJ_{ab}+\calF^a \JJ_a +\calG^a \KK_a\right)\nonumber\\
&+(\varepsilon_1\ast)\calH \DD +(\varepsilon_2\ast) \calF^I \TT_I +(\varepsilon_3\ast)\calF \ZZ\nonumber\\
&+ \QQb (-i\varepsilon_\psi\gf ) \calX + \calXbar(-i\varepsilon_\psi\gf ) \QQ\,.\label{dualop}
\end{align}
and similarly for $\FF'$. $S$ is a symmetry breaking operator that allows to obtain general relativity as the gravity action, see appendix \ref{representationApp}. The parameters $\varepsilon_s$, $\varepsilon_1$, $\varepsilon_2$, $\varepsilon_3$ and $\varepsilon_\psi$ can take values $+1$ or $-1$ only, so that $\circledast^2 = -1$ and therefore the correct sign of the kinetic terms of bosons is ensured. For the present paper we will fix $\varepsilon_s = +1 = \varepsilon_1 = \varepsilon_2$ and $\varepsilon_3 = -1$.

Demanding the diagonal symmetry group to be $su(2,2|5)$ means that not all the fields in $\AAA$ and $\AAA'$ are independent,
\begin{align}
 \omega'^{ab} = &\omega^{ab}\,,\label{diag1}\\
 f'^a = &f^a\,,\\
 g'^a = &g^a\,,\\
 h' = &h\,,\\
 A' = &A\,,\label{diag-2}
\end{align}
and
\begin{equation}
 \left. A'^X \right|_{X=I} = A^I\,.\label{diag-1}
\end{equation}
Therefore (\ref{action}) has only two adjustable parameters in front of the unprimed and primed contributions, $\xi$ and $\xi'$ respectively. The present model is a unified model in the sense that the ratios of \emph{all} the couplings are determined from the superconformal algebra and the way in which the embedding of quarks and leptons is implemented. The weights of the unprimed and primed sector are not determined a priori though, and they can be treated as phenomenological parameters. These parameters define two dimensional parameter space of allowed relative strengths for the gauge couplings $g^{(SU(5))}$ and $g^{(U(1))}$. The relations between gauge couplings will not be arbitrary though as they will be consequence of the superconformal algebra and the particular embedding of bosons and fermions. We discuss this in details in section \ref{gaugecouplingssection}

\section{Gauge couplings}\label{gaugecouplingssection}

In order to obtain an explicit expansion of (\ref{action}), we use the invariant traces defined in the superalgebra, given in appendix \ref{representationApp}, see eqs. (\ref{trace1}) - (\ref{trace-1}) and (\ref{strace1}) - (\ref{strace-1}). After imposing the diagonal symmetry (\ref{diag1})-(\ref{diag-2}),
\begin{align}
\calL =& \frac{1}{4}\varepsilon_s(\xi+\xi')\epsilon_{abcd}\calF^{ab}\calF^{cd}-\varepsilon_1(\xi+\xi')\calH\ast\calH\nonumber\\
&-\frac{1}{2}\varepsilon_2\left(\xi \calF^I\ast \calF^I +\xi'(n-2)\calF'^X\ast \calF'^X\right) \nonumber \\
&-4\varepsilon_{3}\left[\xi(4/n-1)+\xi'(4/d_n-1)\right]\calF \ast \calF\,, \nonumber\\
&-2i\varepsilon_{\psi}\overline{\calX}\gamma_5\calX-\frac{i}{2}\varepsilon_\chi \overline{\calY} \gamma_5 \calY\,.\label{lagrangian}
\end{align}
Here we have restored the label $n$ of the superalgebra $su(2,2|n)$ in order to draw general conclusions about the validity of the model. In the bosonic sector, $\psi_i = 0 = \chi_{ij}$, and imposing the diagonal symmetry along the $su(n)$ generators as well, yields
\begin{align}
\calL_\text{b} =& \frac{1}{4}\varepsilon_s(\xi+\xi')\epsilon_{abcd}\calR^{ab}\calR^{cd}-\varepsilon_1(\xi+\xi')H \ast H\nonumber\\
&-\frac{1}{2}\varepsilon_2\left(\xi +\xi'(n-2) \right)F^I \ast F^I \nonumber \\ &-4\varepsilon_{3}\left[\xi(4/n-1)+\xi'(4/d_n-1)\right] F \ast F \nonumber \\
& -\frac{(n-2)}{2}\varepsilon_2\xi'\left[2F^I\ast F_{1}^I+F_1^I\ast F_{1}^I+F^{\widetilde{X}}\ast F^{\widetilde{X}}\right]\,,
\end{align}
where
\begin{align}
F^{I}&=dA^I+ \frac{1}{2}f_{JK}{}^{I}A^J A^K\,,\\
F_{1}^{I}&=f_{J\widetilde{Y}}{}^{I}A^J A^{\widetilde{Y}}+\frac{1}{2}f_{\widetilde{Y}\widetilde{Z}}{}^{I}A^{\widetilde{Y}} A^{\widetilde{Z}}\,,\label{F1Xtilde}\\
F^{\widetilde{X}}&=dA^{\widetilde{X}}+\frac{1}{2}\left(f_{JK}{}^{\widetilde{X}}A^{J}A^{K}+2f_{J\widetilde{Y}}{}^{\widetilde{X}}A^{J}A^{\widetilde{Y}}+f_{\widetilde{Y}\widetilde{Z}}{}^{\widetilde{X}}A^{\widetilde{Y}}A^{\widetilde{Z}}\right)\,.\label{FXtilde}
\end{align}
In (\ref{F1Xtilde}) and (\ref{FXtilde}) we removed the primed to the $A'^{\widetilde{X}}$ fields to simplify the notation.

Let us now discuss the values of the coupling constants implied by the model. The gravity theory, in the sector defined by
\begin{equation}\label{auxiliaryfields}
 f^a = \rho e^a\,, \qquad g^a = \sigma e^a\,, \quad |\rho| > |\sigma|
\end{equation}
corresponds to general relativity with cosmological constant
\begin{equation}
\frac{1}{16\pi G_N}\int \epsilon_{abcd}\left( \frac{1}{2}R^{ab} e^c e^d -\frac{\Lambda}{12}e^a e^b e^c e^d \right) \subset \mathcal{S}\,,
\end{equation}
under the the following identification of constants
\begin{equation}\label{GNandCC}
 \frac{1}{16\pi G_N}= (\rho^2-\sigma^2)(\xi+\xi')\,, \qquad \Lambda=-3(\rho^2-\sigma^2)=-\frac{3}{2(\xi+\xi')} M_P^2\,.
\end{equation}

\begin{figure}[ht]
  \centering
  \includegraphics[width=.49\linewidth]{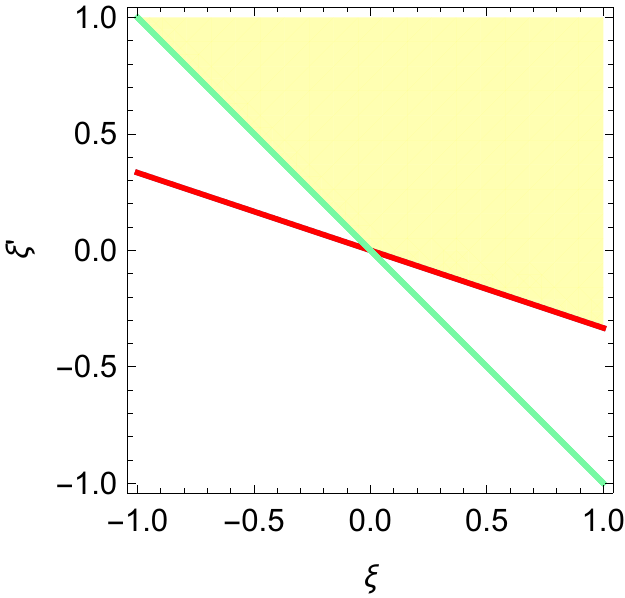}
  \includegraphics[width=.49\linewidth]{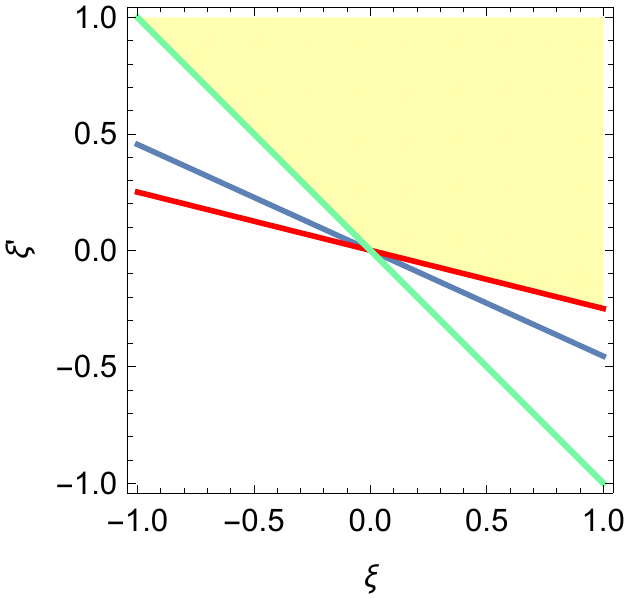}\\
\caption{Allowed regions of the $(\xi,\xi')$ plane, The left panel corresponds to the case $n=5$. The right panel is the plot for $n=6$, which illustrates the qualitative behavior for $n>5$. In green, blue and red are the null conditions for eqs. (\ref{cond1}), (\ref{cond2}) and (\ref{cond3}) respectively. Conditions (\ref{cond2}) and (\ref{cond3}) are degenerate for $n=5$.}
\label{allowedregionsfig}
\end{figure}

The positivity of the kinetic terms for the gauge fields requires
\begin{align}
&\xi+\xi' > 0\,,\label{cond1}\\
&\xi(4/n-1)+\xi'(4/d_n-1) < 0 \,,\label{cond2}\\
&\xi +\xi'(n-2) >0 \,.\label{cond3}
\end{align}
Such conditions are satisfied in a two dimensional region of the parameter space for any $n \geq 5$, see fig. \ref{allowedregionsfig}

The values of the coupling constants are defined with respect to canonically normalized gauge fields,
\begin{equation}
 - a F\ast F = -\half F^ \text{phys} \ast F^ \text{phys}\,,
\end{equation}
where the physical gauge potential is $A^ \text{phys}=\sqrt{2a}A$. Assuming the covariant derivative on the spinor to be of the form
\begin{equation}\label{Dgauge}
 D = d-i g_0 \ \rho(T_r)A^r\,,
\end{equation}
where $\rho(T_r)$ stands for the corresponding representation of the generators $T_r$. Then, the physical values of the coupling constants, $g=g_0/\sqrt{2a}$ can be inferred to be
\begin{align}
&g_{(SU(n))} = g_{(SU(d_n))} = \frac{1}{\sqrt{\xi +\xi'(n-2)}} \,,\label{g1}\\
&g^\text{(rank 1)}_{(U(1))}=\frac{4/n-1}{\sqrt{-8(\xi(4/n-1)+\xi'(4/d_n-1))}}\,,\label{g2}\\
&g^\text{(rank 2)}_{(U(1))}=\frac{4/d_n-1}{\sqrt{-8(\xi(4/n-1)+\xi'(4/d_n-1))}}\label{g3}\,.
\end{align}
The fact that $g_{(SU(n))} = g_{(SU(d_n))}$ allows for a natural embedding of the representations in the GG model (\ref{repGG}) and it will also ensure gauge coupling unification. To simplify the notation we will denote such coupling constants by $g$. The normalization of the gauge field $h$ gives $g^{(U_\DD(1))} = 1/\sqrt{2(\xi+\xi')}$, however, careful evaluation of the fermion kinetic terms in the last two terms of (\ref{lagrangian}), shows that such coupling cancels out and therefore the field $h$ becomes a hidden photon \cite{Arias:2012az}. This is in fact necessary for physical consistency of the model since the transformations generated by $\DD$ are non compact.

\subsection{Fermionic sector}

The fermion kinetic terms are contained in the last two terms of (\ref{lagrangian}),
\begin{equation}
 -2i\calXbar\gamma_5\calX-\frac{i}{2} \calYbar \gamma_5 \calY\,.
\end{equation}
The fact that such a terms contain the right expressions for the kinetic terms of $\psi$ and $\chi$ can be seen as a consequence of a few assumptions:
\begin{enumerate}[(a)]
 \item \label{(a)} The fundamental fermionic fields are in the spin 1/2 representation of the fermionic field in the gauge potential, $\Psi = \se \psi$ and $X = \se \chi$, that are in the kernel of the the spin 3/2 projector, $P_{(3/2)}\Psi =0$ and $P_{(3/2)}X =0$.
 \item The fermionic components of the curvature are the exterior covariant derivatives of $\Psi$ and $X$ with the connection for the corresponding bosonic gauge groups $\calX = D_{G_1} \psi$ and $\calY = D_{G_2} \psi$, where $G_1 = SU(2,2) \times SU(n) \times U(1)$ and $G_2 = SU(2,2) \times SU(d_n) \times U(1)$ respectively.
 \item The dual operator acting on the fermionic components of the curvature is a grading operator of the bosonic gauge symmetries that splits the symmetry between broken an unbroken symmetries. Therefore the covariant derivatives in $\calXbar \lD_\Omega\circledast D_\Omega \calX$ split naturally into $D_\Omega=D_+ + \Omega_-$, where $\Omega=\Omega_+ + \Omega_-$ (for more details see \cite{Alvarez:2021zsw}).
 \item \label{(d)}The fields in $\Omega_-$ are dual to generators that anticommute with $\circledast$, in the present case
 \begin{equation}
  \Omega_-= \half f^a \gamma_a + \half g^a \tgamma_a\,.
 \end{equation}
Furthermore, these fields are related to the orthonormal frames by (\ref{auxiliaryfields}). 
\end{enumerate}
With these assumptions, it can be seen that the terms containing second order derivatives of the fermionic fields are boundary terms, while the first order terms give rise to the gauge covariant kinetic terms. The resulting Lagrangian for the spinor $\psi$ with a connection $\calW \in so(1,3) \times su(n) \times u(1)$ is
\begin{align}
   \frac{\calL(\psi)}{\xi}=& -2i\psibar(\lD_\calW\se\gf\Omega_{-}\se-\se\Omega_{-}\gf\se D_\calW)\psi\nonumber\\
   & -2i\psibar(\sT\gf\Omega_{-}\se+\se\Omega_{-}\gf\sT)\psi\nonumber\\
   %& +2i f^a g_a \psibar\se\se\psi \nonumber\\
   & +2i\psibar\gf\se\left[R^{ab}\Sigma_{ab}-2i(4/n-1)F\right]\se\psi\nonumber\\
   & +2id\left[\psibar\gf\se D_{+}(\se\psi)\right]\,, \label{Lpsi}
\end{align}
where
\begin{equation}
 \Omega_-=\se (\alpha P_R+\beta P_L)\equiv \se \Pi(\alpha,\beta)\,,
\end{equation}
and $P_R$ ($P_L$) is the chiral right-handed projector (left-handed projector). Note here that the term $(1/2) h \DD$ in (\ref{Lpsi}) cancels out from $\calW$ thanks to the fact that $\JJ_a$ and $\KK_a$ anticommute with $\DD$. The constants $\alpha$ and $\beta$ are linear combinations of $\rho$ and $\sigma$
\begin{equation}
 \alpha=\frac{\rho +\sigma}{2}\,, \quad \beta=\frac{\rho -\sigma}{2}\,.
\end{equation}

Similarly for the spinor $\chi$ with a connection $\calW' \in so(1,3) \times su(d_n) \times u(1)$,

\begin{align}
\frac{\mathcal{L}(\chi)}{\xi'}=&
 -\frac{i}{2}\chibar(\lD_{\calW'}\se\gf\Omega_{-}\se-\se\Omega_-\gf\se D_{\calW'})\chi \nonumber\\
 & -\frac{i}{2}\chibar(\sT\gf\Omega_-\se+\se\Omega_-\gf\sT)\chi \nonumber\\
%&+\frac{i}{2} f^{a} g_{a} \chibar\se\se\chi \nonumber\\
& +\frac{i}{2}\chibar\gf\se\left[ R^{ab}\Sigma_{ab}-2i(4/d_n-1) F\right]\se\chi\,,\nonumber\\
& +\frac{i}{2}d[\chibar\gf\se D_+(\se \chi)]\,,\label{Lchi}
\end{align}

The first lines in (\ref{Lpsi}) and (\ref{Lchi}) give the gauge covariant kinetic terms
\begin{align}
  \frac{\calL_K(\psi)+\calL_K(\chi)}{|e|d^{4}x} =& 12 \xi \psibar\left[\alpha P_R(\overleftarrow{\slashed{\nabla}}-\slashed{\nabla})P_L+\beta P_L(\overleftarrow{\slashed{\nabla}}-\slashed{\nabla})P_R\right]\psi \nonumber\\
  &+3 \xi' \chibar\left[\alpha P_R(\overleftarrow{\slashed{\nabla}}{}'-\slashed{\nabla}')P_L+\beta P_L(\overleftarrow{\slashed{\nabla}}{}'-\slashed{\nabla}')P_R\right]\chi\,. \label{gaugekineticfermion0} 
\end{align}
Hence, the physical Weyl spinors are defined by
\begin{align}
 \psi_L^{\text{phys}}=\sqrt{24\alpha\xi} \psi_L\,, \quad \psi_R^\text{phys}=\sqrt{24\beta\xi} \psi_R\,,\\
 \psibar_L^{\text{phys}}=\sqrt{24\alpha\xi} \psibar_L\,, \quad \psibar_R^{\text{phys}}=\sqrt{24\beta\xi} \psibar_R\,,
\end{align}
and
\begin{align}
 \chi_L^{\text{phys}}=\sqrt{6\alpha\xi'} \chi_L\,, \quad \chi_R^\text{phys}=\sqrt{6\beta\xi'} \chi_R\,,\\
 \chibar_L^{\text{phys}}=\sqrt{6\alpha\xi'} \chibar_L\,, \quad \chibar_R^{\text{phys}}=\sqrt{6\beta\xi'} \chibar_R\,.
\end{align}

In order to reproduce a chiral Lagrangian from (\ref{gaugekineticfermion0}) there are two possible scenarios. Firstly, one could assume the spinors in the gauge connection to be chiral,
\begin{equation}
 P_R \psi = 0 \,, \quad P_R \chi =0\,,
\end{equation}
then

\begin{equation}\label{gaugekineticfermion}
  \frac{\calL_K(\psi)+\calL_K(\chi)}{|e|d^{4}x} = \half\psibar_L^{\text{phys}} (\overleftarrow{\slashed{\nabla}}-\slashed{\nabla})\psi_L^{\text{phys}} +\half\chibar_L^{\text{phys}} (\overleftarrow{\slashed{\nabla}}{}'-\slashed{\nabla}')\chi_L^{\text{phys}} \,,
\end{equation}
where
\begin{align}
 \psi_L^{\text{phys}}=\sqrt{24\alpha\xi} \psi_L\,, \\
 \psibar_L^{\text{phys}}=\sqrt{24\alpha\xi} \psibar_L\,,
\end{align}
and
\begin{align}
 \chi_L^{\text{phys}}=\sqrt{6\alpha\xi'} \chi_L\,,\\
 \chibar_L^{\text{phys}}=\sqrt{6\alpha\xi'} \chibar_L\,.
\end{align}

Another possibility is to assume that $\beta \rightarrow 0$, then the gauge covariant kinetic term takes the same form, (\ref{gaugekineticfermion}) and the physical Weyl spinors are
\begin{align}
 \psi_L^{\text{phys}}=\sqrt{24\alpha\xi} \psi_L\,, \quad \psi_R^\text{phys}=0\,,\\
 \psibar_L^{\text{phys}}=\sqrt{24\alpha\xi} \psibar_L\,, \quad \psibar_R^{\text{phys}}=0\,,
\end{align}
and
\begin{align}
 \chi_L^{\text{phys}}=\sqrt{6\alpha\xi'} \chi_L\,, \quad \chi_R^\text{phys}=0\,,\\
 \chibar_L^{\text{phys}}=\sqrt{6\alpha\xi'} \chibar_L\,, \quad \chibar_R^{\text{phys}}=0\,,
\end{align}

It can be seen that $\psi_R^\text{phys}=0$ and $\chi_R^\text{phys}=0$ is a consistent choice such that the bilinears that mix chiralities are well defined
\begin{align}
 \phi =& \frac{1}{24\sqrt{\alpha\beta}}\phi^\text{phys}\,,\label{scalarphysical}\\
 \phi_5 =& \frac{1}{24\sqrt{\alpha\beta}}\phi_5^\text{phys}\label{pseudoscalarphysical}\\
 \phi_{ab} =& \frac{1}{24\sqrt{\alpha\beta}}\phi_{ab}^\text{phys}\,,\label{tensorphysical}
\end{align}
where $\phi_X=\psibar \gamma_X \psi$. It remains an interesting question to address in the future to determine whether or not there are consistent alternatives to $\psi_R^\text{phys}=0$ and $\chi_R^\text{phys}=0$ when $\beta \rightarrow 0$. 

Let us now show the connection to the $SU(5)$ GG model. The covariant derivatives in (\ref{gaugekineticfermion}) expand as
\begin{align}
 \slashed{\nabla}\psi_L^{\text{phys}} = & \slashed{\nabla}_{su(5)}\psi_L^{\text{phys}}-ig^\text{(rank 1)}_{(U(1))}\slashed{A} \psi_L^{\text{phys}}\,,\label{covderpsi}\\
 \slashed{\nabla}'\chi_L^{\text{phys}} = & \slashed{\nabla}_{su(5)}\chi_L^{\text{phys}}-ig \slashed{A}^{\tilde{X}}t_{\tilde{X}}\chi_L^{\text{phys}}-ig^\text{(rank 2)}_{(U(1))}\slashed{A} \chi_L^{\text{phys}}\,,\label{covderchi}
\end{align}
where the coupling constants are given in (\ref{g1})-(\ref{g3}). The generators $t_X$ in (\ref{covderchi}) carry a rank 2 antisymmetric representation of $su(10)$, where $t_X$ with $X=I=1,\cdots,24$ form a $su(5)$ algebra. The index $\widetilde{X}$ runs for the range $=25,\cdots,99$, see reference \cite{Alvarez:2021lda} or Appendix \ref{rank2algebraApp} for more details.

From the first terms in (\ref{covderpsi}) and (\ref{covderchi}), the couplings between fermions and the gauge sector of the GG model can be recovered. We also get new gauge couplings not present in the GG model coming from the last terms in (\ref{covderpsi}) and (\ref{covderchi}).

\subsection{Gauge potentials}

In the GG model the gauge bosons of $SU(5)$ transform in the adjoint representation that decomposes as
\begin{equation}
 \mathbf{24} = (\mathbf{8},\mathbf{1},0) + (\mathbf{1},\mathbf{3},0) + (\mathbf{1},\mathbf{1},0) +(\mathbf{3},\mathbf{2}^\ast,-\tfrac{5}{6}) + (\mathbf{3}^\ast,\mathbf{2},\tfrac{5}{6})\,.
\end{equation}
The first three representations are the gluons, the $W$-bosons and the $B$-boson. The last two representations carry both color and flavor and are associated to colored and electrically charged gauge bosons $\text{X}_i$ and $\text{Y}_i$ (here $i \in SU(3)$) that form an $SU(2)$ doublet. The hypercharge and electric charge generators are
\begin{align}
 Y =& \text{diag} (-1/3,-1/3,-1/3,1/2,1/2)\,,\\
 Q =& \text{diag} (-1/3,-1/3,-1/3,1,0)\,,
\end{align}
when embedded in the $SU(5)$ block of the superalgbera. When $Q$ is embedded in the superalgebra, the charge assignments of the fermions can be verified using 
\begin{equation}
    [Q,\upsilon] = q(Q) \upsilon \,,
\end{equation}
where $\upsilon$ is 
\begin{equation}
    \Psi = \QQ_i\psi^i \quad \text{or} \quad \Phi = \QQb^{ij}\chi_{ij}\,.
\end{equation}

It is convenient to visualize the full $su(5)$ matrix in the following form
\begin{equation}
A^I (-i (t_I)_i{}^j)=\frac{1}{\sqrt{2}} A_i{}^j\,, 
\end{equation}
where
\begin{equation}
 A_i{}^j = \left(\begin{array}{ccc|cc}
            G_1{}^1 -\frac{2B}{\sqrt{30}} & G_1{}^2 & G_1{}^3 & \overline{\text{X}}_1 & \overline{\text{Y}}_1 \\ 
            G_2{}^1 & G_2{}^2 -\frac{2B}{\sqrt{30}} & G_2{}^3 & \overline{\text{X}}_2 & \overline{\text{Y}}_2 \\
            G_3{}^1 & G_3{}^2 & G_3{}^3 -\frac{2B}{\sqrt{30}} & \overline{\text{X}}_3 & \overline{\text{Y}}_3 \\ \hline
            \text{X}^1 & \text{X}^2 & \text{X}^3 & \frac{W^3}{\sqrt{2}} + \frac{3B}{\sqrt{30}} & W^+\\
            \text{Y}^1 & \text{Y}^2 & \text{Y}^3 & W^- & -\frac{W^3}{\sqrt{2}} + \frac{3B}{\sqrt{30}} 
           \end{array}\right)\,.\label{su(5)matrix}
\end{equation}
The electric charges of particles X and Y can also be verified using their embedding in the superconformal algebra, obtaining $q(\text{X})=4/3$ and $q(\text{Y})=1/3$.

The gauge bosons $A^\wtX$ generate gauge kinetic couplings that are not present in the GG model (see (\ref{gaugenew})). Before carrying out the actual computation it is convenient to split the $SU(10)$ block into the $SU(5)$ block of the GG model and to also split the $U(1)$ blocks of the superconformal algebra,
\begin{align}
 \psibar_L \slashed{\nabla}\psi_L + \chibar_L \slashed{\nabla}'\chi_L = & \psibar_L \slashed{\nabla}_{su(5)}\psi_L + \chibar_L \slashed{\nabla}_{su(5)}\chi_L - ig \chibar_L\slashed{A}^{\wtX}t_{\wtX}\chi_L \nonumber\\
 & - i\psibar_L g^\text{(rank 1)}_{(U(1))}\slashed{A} \psi_L - i\chibar_L g^\text{(rank 2)}_{(U(1))}\slashed{A} \chi_L \,,\label{gaugekineticsectornabla}
\end{align}
where we dropped the label of the physical spinor to simplify the notation. The gauge coupling in the first two terms of (\ref{gaugekineticsectornabla}) has the standard form of the GG model,
\begin{align}
\mathcal{L}^\text{GG}_{G} = & g \left[\overline{u}\slashed{G}u+\overline{d}\slashed{G}d \right]
    + g \left[\left(\begin{smallmatrix}
              \overline{u} & \overline{d} \\
              \end{smallmatrix}\right)_L \slashed{W} \left(\begin{smallmatrix}
                                                     \overline{u} \\
                                                     \overline{d} \\
                                                     \end{smallmatrix}\right)_L
    +\left(\begin{smallmatrix}
           \overline{e}^- & \overline{\nu}_e \\
           \end{smallmatrix}\right)_L\slashed{W} \left(\begin{smallmatrix}
                                                       \overline{e}^- \\
                                                       \overline{\nu}_e \\
                                                       \end{smallmatrix}\right)_L\right]\nonumber\\
& + g \sqrt{\frac{3}{5}}\left[-\frac{1}{2}\left(\overline{e}^-_L\slashed{B}e^-_L +\overline{\nu}_{eL}\slashed{B}\nu_{eL}\right) +\frac{1}{6}\left(\overline{u}_L\slashed{B}u_L+\overline{d}_L\slashed{B}d_L \right)\right.\nonumber\\
&\left.+\frac{2}{3}\overline{u}_R\slashed{B}u_R-\frac{1}{3}\overline{d}_R\slashed{B}d_R -\overline{e}^{+}_R\slashed{B} e^{+}_R\right]\nonumber\\
& + \frac{g}{\sqrt{2}}\left[\overline{d}^i_R\slashed{\overline{X}}_i e^{+}_R + \overline{d}^i_{L}\slashed{\overline{X}}_ie^{+}_{L}+\epsilon^{ijk}\overline{u}^c_{iL}\slashed{\overline{X}}_j u_{kL}\right]+\text{H.C.} \nonumber\\
& + \frac{g}{\sqrt{2}}\left[-\overline{d}^i_R\slashed{\overline{Y}}_i \nu^c_R - \overline{u}^i_{L}\slashed{\overline{Y}}_ie^{+}_{L}+\epsilon^{ijk}\overline{u}^c_{iL}\slashed{\overline{Y}}_j d_{kL}\right]+\text{H.C.}\label{GGgauge}
\end{align}
where
\begin{align}
 \slashed{G} =& \gamma^\mu G^I_\mu \left(-i \frac{\lambda_I}{2}\right)\,, \quad I \in SU(3)\label{gellmann}\\
 \slashed{W} =& \gamma^\mu W^I_\mu \left(-i \frac{\tau_I}{2}\right)\,, \quad I \in SU(2)\label{pauli}\\
 \slashed{B} =& \gamma^\mu B_\mu\,,\\
 \slashed{\overline{X}}_i =& \gamma^\mu \overline{\text{X}}_{\mu i}\,, \quad \slashed{\overline{Y}}_i = \gamma^\mu \overline{\text{Y}}_{\mu i}\,,
\end{align}
and $\lambda^I$ and $\tau^I$ are Gell-Mann and Pauli matrices respectively.\footnote{The extra minus sign in the definitions (\ref{gellmann}) and (\ref{pauli}) w.r.t. reference \cite{Langacker:1980js} comes from the fact that we are using antihermitian generators $\lambda_I=i\lambda^\text{(Gell-Mann)}_I$, etc., which is motivated by the conventions in supergravity.} One of the beautiful aspects of the GG model is the correct assignment of fractional quark charges, and that is also valid in the present model. 

From (\ref{gaugekineticsectornabla}) we also get new couplings that are not present in the GG model.
\begin{align}
\mathcal{L}^{\text{new}}_{G} = & -i g^\text{(rank 1)}_{(U(1))} \left[\overline{d}^c_L \slashed{A} d^c_L+ \overline{e}^-_L \slashed{A}e^-_L + \overline{\nu}_{eL} \slashed{A}\nu_{eL}\right]\\
&- i g^\text{(rank 2)}_{(U(1))} \left[\overline{u}^c_L \slashed{A}u^c_L+\overline{u}_L \slashed{A}u_L +\overline{d}_L \slashed{A}d_L+\overline{e}^+_L \slashed{A}e^+_L\right]\\
&+  i g \chibar_L \slashed{A}^\wtX t_\wtX \chi_L\,,\label{gaugenew}
\end{align}
The last line in (\ref{gaugenew}) has the interactions mediated by the $SU(10)$ gauge bosons that are not in the $SU(5)$ block (\ref{su(5)matrix}). The matrix elements of $t_\wtX$ have been defined in \cite{Alvarez:2021lda} in no particular basis, as obtained by a Gramsh-Schmidt procedure. From the point of view of phenomenological applications it is useful to consider the following decomposition of the adjoint representation of $SU(10)$
\begin{equation}
 \mathbf{99} = (\mathbf{24},\mathbf{1},0) + (\mathbf{1},\mathbf{24},0) + (\mathbf{1},\mathbf{1},0) + (\mathbf{5},\mathbf{5}^\ast,-y') + (\mathbf{5}^\ast,\mathbf{5},y')\,.
\end{equation}
The fields in (\ref{su(5)matrix}) are uncharged with respect to the new $(\mathbf{1},\mathbf{24},0)$ representation. There is also a quintet-quintet set of bosons charged respect to the $(\mathbf{24},\mathbf{1},0)$ and the $(\mathbf{1},\mathbf{24},0)$ representation. In addition, there is a new hypercharge-like operator associated to a new $B$-boson.

\subsection{Comments on anomalies}\label{annomalies}

As it happens in the GG model, anomaly cancellation of the cubic $SU(3)$, $SU(2)$ and hypercharge are preserved and received a rationale within the $SU(5)$ block \cite{Langacker:1980js}. What seems to be problematic now are the extra $U(1)_Z$ groups produced by the generators along the diagonal in the superconformal algebra.

The $U(1)_Z$ charges are computed according to
\begin{equation}
[Z^{\text{(rank 1)}},\Psi]= x_1 \Psi\,,  \quad [Z^{\text{(rank 2)}},\text{X}]= x_2 \text{X}\,,
\end{equation}
where
\begin{equation}
 x_1 = \left(\frac{- i z_1}{5}\right)\,, \quad x_2 = \left(\frac{3 i z_2}{5}\right)\,,
\end{equation}
and $z_1$ and $z_2$ are the overall constants of the $U(1)_Z$ generator in the fundamental and bi-fundamental representation, (\ref{Z1gen}) and (\ref{Z2gen}), respectively. In the following table we have a list of the quarks and leptons, hypercharges and $Z$-charges. 
\begin{table}[h]
\begin{center}
\begin{tabular}{l|llll}
  & $SU(3)$ & $SU(2)$ & $Y$ & $Z$ \\ \hline
 $Q_L$ & $\mathbf{3}$ & $\mathbf{2}$ & $+1/6$ & $x_2$ \\
 $L_L$ & $\mathbf{1}$ & $\mathbf{2}$ & $-1/2$ & $x_1$ \\ \hline
 $u_R$ & $\mathbf{3}$ & $\mathbf{1}$ & $+2/3$ & $x_2$ \\
 $d_R$ & $\mathbf{3}$ & $\mathbf{1}$ & $-1/3$ & $x_1$ \\
 $e_R$ & $\mathbf{1}$ & $\mathbf{1}$ & $-1$ & $x_2$
\end{tabular}
\end{center}
\end{table}
Besides the cancellation of the mixed $(Y \ SU(3)^2)$, $(Y \ SU(2)^2)$ anomalies, and of the cancellation of the cubic anomaly $Y^3$, we observe that the cubic $Z^3$ and the mixed anomalies $(Z \ SU(3)^2)$, $(Z \ SU(3)^2)$, cannot possibly all cancel simultaneously in the present model:
\begin{align}
 A_Z =& 6 x_2^3+2 x_1^3-\left( 3 x_2^3+3x_1^3+ x_2^3\right)\,,\\
 A_{SU(2)} =& 3 x_2 + x_1\,,\\
 A_{SU(3)} =& 2 x_2 - \left( x_2+x_1\right)\,.
\end{align}
Therefore, in order to avoid the anomaly induced by the $Z$ symmetry we would have to invoke anomaly cancellation either between families, or by considering of left-right symmetric models such as the Pati-Salam, or by including new matter content \cite{Bilal:2008qx}.

\section{Conclusions}\label{conclu}

In the present paper we have constructed a grand unified model using the superconformal algebra and matter in the adjoint representation of supersymmetry. The model reproduces the main features, charges and families of the Georgi-Glashow model, starting from an unconventional form of supersymmetry that includes gravity from
the start \cite{Alvarez:2013tga,Alvarez:2021zsw}. In this construction, local SUSY is not an invariance of the action but could be feature the ground state \cite{Alvarez:2021qbu}.
The model predicts a hidden sector $(\mathbf{1},\mathbf{24},0)  + (\mathbf{5},\mathbf{5}^\ast,-y') + (\mathbf{5}^\ast,\mathbf{5},y')$ sector and an anomalous $U(1)_Z$.

A very interesting issue to address as a future work is the computation of the anomaly from the point of view of the embedding of the superconformal algebra, in terms of the symmetric tensor 
\begin{eqnarray}
d^{MNL}=\half \langle T^M \{ T^N,T^L\} \rangle\,,
\end{eqnarray}
where $\langle X \rangle$ stands for the supertrace of $X$, and also including the gradding associated to the operator $\circledast$ \cite{Wan:2019gqr,Davighi:2019rcd}. Such study could shed light into the question of existence of representations of the superconformal algebra that, perhaps adding new matter fields, could be anomaly free.

Another aspect of the model that we leave for a future study is the introduction of spontaneous symmetry breaking with a Higgs potential.

%%%%%%%%%%%%%%%%%%%%%%%%%%%%%%%%%%%%%%
\section*{Acknowlegements}
%%%%%%%%%%%%%%%%%%%%%%%%%%%%%%

We thank Biancha Cerchiai, Laura Andrianopoli, Mario Trigiante and Lucrezia Ravera for useful discussions. P. A. acknowledges MINEDUC-UA project ANT 1755 and Semillero de Investigación project SEM18-02 from Universidad de Antofagasta, Chile. This research has been partially funded by Fondecyt Grant 1220862.

\begin{appendices}

\section{Fundamental representation of $SU(2,2|n)$}\label{representationApp}

Let us consider the following representation of $SU(2,2|n)$
\begin{eqnarray}
&\JJ_a =\left[\begin{array}{c|c}
\frac{s}{2}\gamma_a &  0_{4\times n}\\[0.5em] \hline
0_{n\times4} & 0_{n\times n} \\
\end{array}\right]\,, \quad \text{or} \quad (\JJ_a)^A_{\ B}=\frac{s}{2}(\gamma_a)^\alpha_{\ \beta}\delta^A_{\ \alpha} \delta^\beta_{\ B}=\frac{s}{2}(\gamma_a)^A_{\ B}\,,&\\
&\JJ_{ab} =\left[\begin{array}{c|c}
\frac{1}{4}[\gamma_a,\gamma_b] &  0_{4\times n }\\[0.5em] \hline
0_{n\times4} & 0_{n\times n} \\
\end{array}\right]\,, \quad \text{or} \quad (\JJ_{ab})^A_{\ B}=\frac{1}{4}[\gamma_a,\gamma_b]^A_{\ B}=(\Sigma_{ab})^A_{\ B}\,,&\\
&\KK_a =\left[\begin{array}{c|c}
\frac{1}{2}\tilde{\gamma}_a &  0_{4\times n}\\[0.5em] \hline
0_{n\times4} & 0_{n\times n} \\
\end{array}\right]\,, \quad \text{or} \quad (\KK_a)^A_{\ B}=\frac{1}{2}(\tilde{\gamma}_a)^A_{\ B}\,,&\\
&\DD =\left[\begin{array}{c|c}
\frac{1}{2}\gamma_5 &  0_{4\times n}\\[0.5em] \hline
0_{n \times4} & 0_{n\times n} \\
\end{array}\right]\,, \quad \text{or} \quad (\DD)^A_{\ B}=\frac{1}{2}(\gamma_5)^A_{\ B}\,,&\\
&\TT_{I} =\left[\begin{array}{c|c}
0_{4\times4} &  0_{4\times n}\\[0.5em] \hline
0_{n \times4} & \frac{i}{2}\lambda_{I}^{t} \\
\end{array}\right]\,, \quad \text{or} \quad (\TT_I)^A_{\ B}=\frac{i}{2}(\lambda_I^{t})_{\ B}^{A} \,,&\\
&(\QQ^\alpha_i)^A_{\ B}=\left[\begin{array}{c|c}
0_{4\times4} & 0_{4\times n}\\ [0.5em] \hline
\delta^A_{i} \delta^\alpha_B & 0_{n\times n}
\end{array}\right]=\delta^A_{i} \delta^\alpha_B\,,&\\
&(\QQb_\alpha^i)^A_{\ B}=\left[\begin{array}{c|c}
0_{4\times4} & \delta^A_\alpha \delta^i_B\\ [0.5em] \hline
0_{n\times 4} & 0_{n\times n}
\end{array}\right]=\delta^A_\alpha \delta^i_B\,,&\\
&\ZZ^A_{\ B}=z_1\left[\begin{array}{c|c}
i\delta^\alpha_\beta &0_{4\times n}\\ [0.5em] \hline
0_{n\times4} & \frac{4}{n} i\delta^i_j\end{array}\right]=
z_1\left(i\delta^A_\alpha\delta^\alpha_B+\frac{4i}{n}\delta^A_i\delta^i_B\right)\,,\label{Z1gen}&
\end{eqnarray}
where $\gamma_5=i\gamma^0 \gamma^1 \gamma^2 \gamma^3$, $(\gamma_5)^2=\mathds{1}$,
$$\gamma_{abc}=\gamma_{[a}\gamma_b \gamma_{c]}=\frac{1}{3!}\sum_{\Pi(a,b,c)} \text{sign}(\Pi(a,b,c)) \gamma_a \gamma_b \gamma_c=i\epsilon_{abcd}\gamma_5 \gamma^{d}\,.$$
and $$\tilde{\gamma}_a=\frac{i}{3!}\epsilon_{abcd}\gamma^{bcd}=-\gamma_5\gamma_a\,,$$

The $\gamma$-matrices are in a $4\times 4$ spinor-representation ($\alpha, \beta,\cdots$ run from 1 to 4). The indices of the tangent space $a,b=0,1,2,3$. Indices in the adjoint representation of $su(n)$ take values $I,J=1,2,\ldots,n^{2}-1$, and in the fundamental take the values $i,j=1,2,\ldots,n$. The $\gamma$-matrices are endomorphisms and they act on spinors
\begin{equation}
 \psi^\alpha \stackrel{\gamma_a}{\longrightarrow} (\gamma_a)^\alpha_{\ \beta} \psi^\beta\,.
\end{equation}
These $\gamma$-matrices satisfy $\{\gamma^a,\gamma^b\}=2 \eta^{ab}$, where the metric $\eta$ is given by $\eta=\mathrm{diag}(-,+,+,+)$. The spinor indices will be often omitted.

In a similar way the $\lambda$-matrices are also endomorphisms and they act on spinors as
\begin{equation}
 \psi^\alpha_i \stackrel{\lambda_I}{\longrightarrow} (\lambda_I)_i^{\ j}\psi^\alpha_j \,.
\end{equation}
The $\lambda$-matrices satisfy $[\lambda_I,\lambda_J]=f^{IJK} \lambda_K$.

Indices of the representation are $A,B=1,\cdots,n+4$, so we have a $(n+4) \times (n+4) $ representation. We find it convenient to use the splitting $A=(\alpha,i)$. All the possible products that mix tensors from different spaces, like $p^i_{\ A}  q^A_{\ \alpha}$, are trivial. The following relations are understood
\begin{eqnarray}
&(\gamma_a)^A_{\ B}=\delta^A_\alpha (\gamma_a)^\alpha_{\ \beta} \delta^\beta_B\,,&\\
&C_{\alpha A}=C_{\alpha\beta} \delta^\beta_A \,.&
\end{eqnarray}

The generators $\JJ_a$ and $\JJ_{ab}$ form an adS$_4$ algebra,
\begin{equation}
[\JJ_a,\JJ_b]=s^2\JJ_{ab}\,,
\end{equation}
\begin{equation}
[\JJ_a,\JJ_{bc}]=\eta_{ab}\JJ_c-\eta_{ac}\JJ_b\,,
\end{equation}
\begin{equation}
[\JJ_{ab},\JJ_{cd}]=-(\eta_{ac}\JJ_{bd}-\eta_{ad}\JJ_{bc}-\eta_{bc}\JJ_{ad}+\eta_{bd}\JJ_{ac})\,.
\end{equation}
The parameter $s^2$ can take values $s^2=+1,-1$ for anti-de Sitter or de Sitter algebras, respectively.

Among $\DD$ and $\KK_a$ they form the conformal algebra,
\begin{equation}
[\KK_a,\KK_b]=-\JJ_{ab}\,,
\end{equation}
\begin{equation}
[\JJ_a,\KK_{b}]=s\eta_{ab}\DD\,,
\end{equation}
\begin{equation}
[\KK_a,\JJ_{bc}]=\eta_{ab}\KK_c-\eta_{ac}\KK_b\,,
\end{equation}
\begin{equation}
[\DD,\KK_{a}]=-s^{-1}\JJ_a\,,
\end{equation}
\begin{equation}
[\DD,\JJ_{a}]=-s\KK_a\,.
\end{equation}

For the internal generators we have the $su(n)$ algebra
\begin{equation}
 [\TT_I,\TT_J]=f_{IJ}{}^K\TT_K\,,
\end{equation}
and they are anti-hermitian $\TT_I^\dag=-\TT_I$ (also $\ZZ^\dag=-\ZZ$).

Including $\QQ^\alpha_i$ and $\QQb^i_\alpha$ the commutators close in a $su(2,2|n)$ superalgebra
\begin{align}
&[\JJ_a,\QQb_\alpha^i]=\frac{s}{2}\QQb_\beta^i(\gamma_a)^\beta_{\ \alpha}\,, \quad [\JJ_a,\QQ^\alpha_i]=-\frac{s}{2}(\gamma_a)^\alpha_{\ \beta}\QQ^\beta_i\,,\\
&[\JJ_{ab},\QQb_\alpha^i]=\QQb_\beta^i(\Sigma_{ab})^\beta_{\ \alpha}\,, \quad [\JJ_{ab},\QQ^\alpha_i]=-(\Sigma_{ab})^\alpha_{\ \beta}\QQ^\beta_i\,,\\
&[\KK_a,\QQb_\alpha^i]=\frac{1}{2}\QQb_\beta^i(\tilde{\gamma}_a)^\beta_{\ \alpha}\,, \quad [\KK_a,\QQ^\alpha_i]=-\frac{1}{2}(\tilde{\gamma}_a)^\alpha_{\ \beta}\QQ^\beta_i\,,\\
&[\DD,\QQb_\alpha^i]=\frac{1}{2}\QQb_\beta^i(\gamma_5)^\beta_{\ \alpha}\,, \quad [\DD,\QQ^\alpha_i]=-\frac{1}{2}(\gamma_5)^\alpha_{\ \beta}\QQ^\beta_i\,,\\
&[\TT_I,\QQb_\alpha^i]=-\frac{i}{2}\QQb_\alpha^j (\lambda_I)^{\ i}_j\,, \quad [\TT_I,\QQ^\alpha_i]=\frac{i}{2}(\lambda_I)^{\ j}_i\QQ^\alpha_j\,,\\
&[\ZZ,\QQb_\alpha^i]=-iz_1(4/n-1)\QQb_\alpha^i\,, \quad [\ZZ,\QQ^\alpha_i]=iz_1(4/n-1)\QQ^\alpha_i\,,\\
&\{\QQ^\alpha_i,\QQb_\beta^j\}=(\gamma^C)^\alpha{}_\beta \delta^j_i \JJ_C +\delta^\alpha_\beta\left(-i(\lambda_I)_i^{\ j}\TT_I-\frac{i}{4z_1}\delta_i^{j} \ZZ\right)\,,
\end{align}
where we have used the short-hand
\begin{equation}
 (\gamma^C)^\alpha{}_\beta \JJ_C = \left(\frac{1}{2s}(\gamma^a)^\alpha_{\ \beta} \JJ_a-\frac{1}{2}(\Sigma^{ab})^\alpha_{\ \beta} \JJ_{ab}-\frac{1}{2}(\tilde{\gamma}^a)^\alpha_{\ \beta} \KK_a+\frac{1}{2}(\gamma_5)^\alpha_{\ \beta} \DD\right)\,.
\end{equation}

\subsection*{Traces:}
The gradding operator is given by
\begin{equation}
\mathcal{G}^A_{\ B} = \delta^A_{\alpha}\delta^\alpha_{\ B}-\delta^{A}_{i}\delta^i_{\ B}\,,
\end{equation}
which classifies generators in bosonic $B=\{\JJ_a,\JJ_{ab},\KK_a,\DD,\TT_I,\ZZ\}$ or fermionic $F=\{Q^\alpha_i,\QQb^i_\alpha\}$, by $[B,\mathcal{G}]=0=\{F,\mathcal{G}\}$, and squares to one, $\mathcal{G}^2=1$. The gradding operator defines an invariant supertrace
\begin{equation}
\langle G\rangle  \equiv Tr(\mathcal{G} G)=0\,.
\end{equation}
The supertrace has the following properties
\begin{equation}
 \langle B_1 B_2\rangle =\langle B_2 B_1\rangle \,, \quad \langle B F\rangle =\langle F B \rangle \,, \quad \langle F_1 F_2\rangle =-\langle F_2 F_1\rangle \,.
\end{equation}

All generators $G$ in the representation are super-traceless
\begin{equation}
\langle G\rangle=0\,, \quad G=\{\JJ_a,\JJ_{ab},\KK_a,\DD,\TT_I,\ZZ,Q^\alpha_i,\QQb^i_\alpha\}\,.
\end{equation}
The quadratic combinations that give nontrivial traces are
\begin{eqnarray}
&\langle \JJ_a \JJ_b \rangle=s^2\eta_{ab}\,, \qquad \langle \JJ_{ab} \JJ_{cd} \rangle=-(\eta_{ac}\eta_{bd}-\eta_{bc}\eta_{ad})\,,&\label{trace1}\\
& \langle \KK_a \KK_b \rangle=-\eta_{ab}\,, \qquad \langle \DD^2\rangle = +1\,,&\label{trace2}\\
&\langle \TT_I \TT_J \rangle=\frac{1}{2}\delta_{IJ}\,, \qquad \langle \ZZ^2\rangle=4z_1^2(4/n-1)\,,&\\
&\langle \QQ^\alpha_i \QQb^j_\beta\rangle=-\delta^\alpha_\beta \delta^j_i=-\langle \QQb^j_\beta \QQ^\alpha_i\rangle\,.&\label{trace-1}
\end{eqnarray}

\subsection*{$S$-grading operator}

The $S$ operator is essential for the generalized $\circledast$ dual opperator, as in (\ref{dualop}). The nontrivial traces involving $S$ are
\begin{align}
 \langle S \DD \rangle &=2i\varepsilon_s\,, \label{strace1}\\
 \langle S \JJ_{ab}\JJ_{cd}\rangle&=-\varepsilon_s\epsilon_{abcd}=\langle \JJ_{ab}S\JJ_{cd}\rangle\,,\\
 \langle  \JJ_{a}S\KK_{b}\rangle&=-i\varepsilon_s s \eta_{ab}=-\langle \KK_{a}S\JJ_{b}\rangle\,,\\
 \langle \ZZ S \DD \rangle&=-2z_1\varepsilon_s=\langle \DD S \ZZ \rangle\,.\label{strace-1}
\end{align}
These traces are instrumental in producing the usual expressions for the standard kinetic terms in the action.

\section{Bi-fundamental representation of $SU(2,2|n)$}\label{rank2algebraApp}

The generators of the conformal group $SU(2,2)$, $\{ \JJ_C \} = \{ \JJ_a,\JJ_{ab},\KK_a,\DD \}$, embedded in the superalgebra are:
\begin{align}
&\JJ_a =\left[\begin{array}{c|c}
\frac{s}{2}\gamma_a &  0_{4\times d_n}\\[0.5em] \hline
0_{d_n\times4} & 0_{d_n\times d_n} \label{Ja}\\
\end{array}\right]\,, \quad \text{or} \quad (\JJ_a)^A_{\ B}=\frac{s}{2}(\gamma_a)^\alpha_{\ \beta}\delta^A_{\ \alpha} \delta^\beta_{\ B}=\frac{s}{2}(\gamma_a)^A_{\ B}\,,&\\
&\JJ_{ab} =\left[\begin{array}{c|c}
\frac{1}{4}[\gamma_a,\gamma_b] &  0_{4\times d_n}\\[0.5em] \hline
0_{d_n\times4} & 0_{d_n\times d_n} \\
\end{array}\right]\,, \quad \text{or} \quad (\JJ_{ab})^A_{\ B}=\frac{1}{4}[\gamma_a,\gamma_b]^A_{\ B}=(\Sigma_{ab})^A_{\ B}\,,&\\
&\KK_a =\left[\begin{array}{c|c}
\frac{1}{2}\tilde{\gamma}_a &  0_{4\times d_n}\\[0.5em] \hline
0_{d_n\times4} & 0_{d_n\times d_n} \\
\end{array}\right]\,, \quad \text{or} \quad (\KK_a)^A_{\ B}=\frac{1}{2}(\tilde{\gamma}_a)^A_{\ B}\,,&\\
&\DD =\left[\begin{array}{c|c}
\frac{1}{2}\gamma_5 &  0_{4\times d_n}\\[0.5em] \hline
0_{d_n\times4} & 0_{d_n\times d_n} \\
\end{array}\right]\,, \quad \text{or} \quad (\DD)^A_{\ B}=\frac{1}{2}(\gamma_5)^A_{\ B}\,,&\label{D}
\end{align}

\begin{equation}\label{Z2gen}
 \ZZ^A_{\ B}=i z_2 \left(\delta^A_\alpha \delta^\alpha_B+\frac{4}{d_n}\Delta^A_{ij}\Delta^{ij}_B\right)\,.
\end{equation}

The following commutators get modified with respect to the fundamental representation, and they close in $su(2,2|d_n)$:
\begin{equation}
 [\GG_{X'},\GG_{Y'}]= f_{X'Y'}{}^{Z'} \GG_{Z'}\,,
\end{equation}
and
\begin{eqnarray}
&[\GG_{X'},\QQ^\alpha_{ij}]=i(g_{X'})_{ij}{}^{kl}\QQ^\alpha_{kl}\,, \quad [\GG_{X'},\QQb_\alpha^{ij}]=-i\QQb_\alpha^{kl} (g_{X'})_{kl}{}^{ij}\,,&\\
&[\ZZ,\QQ^\alpha_{ij}]=iz_2\left(4/d_n-1\right)\QQ^\alpha_{ij}\,, \quad [\ZZ,\QQb_\alpha^{ij}]=-iz_2\left(4/d_n-1\right)\QQb_\alpha^{ij}\,,&\label{Zcomm}\\
&\{\QQ^\alpha_{ij},\QQb_\beta^{kl}\}=(\gamma^C)^\alpha{}_\beta \Delta_{ij}^{kl} \JJ_C-\left(\frac{4i}{n-2}(g_{X'})_{ij}{}^{kl}\GG_{X'}+\frac{i}{2z_2}\Delta_{ij}^{kl} \ZZ\right) \delta^{\alpha}_{\beta}\,.&\\
\end{eqnarray}

The generators $g_{X'}$ and $v_X$, that belong to the bifundamental rep. of $SU(n)$ and the fundamental rep. of $SU(d_n)$, are related to each other by means of the change of basis matrix defined by
\begin{equation}\label{relationgxvx}
 g_{X'} = C_{X'}{}^X v_X\,.
\end{equation}
We will define the inverse matrix by
\begin{equation}\label{inverserelation}
 v_X = C_{X}{}^{X'} g_{X'}\,,
\end{equation}
which satisfies
\begin{equation}
 C_{X}{}^{X'}C_{X}{}^{Y'}=\frac{1}{n-2}\delta^{X'Y'}\,.
\end{equation}

Let us define the structure constants of $su(d_n)$ in the fundamental representation by
\begin{equation}
 [v_X,v_Y]=i F_{XY}{}^Z v_Z\,.
\end{equation}
Using the matrix $C_{X}{}^{X'}$ and its inverse we can determine the structure constants $f_{X'Y'}{}^{Z'}$, that are defined in the $g_{X'}$ basis
\begin{equation}
 [g_{X'},g_{Y'}]=i f_{X'Y'}{}^{Z'} g_{Z'}\,,
\end{equation}
where
\begin{equation}
f_{X'Y'}{}^{Z'} =C_{X'}{}^X C_{Y'}{}^Y F_{XY}{}^Z C_{Z}{}^{Z'}\,.
\end{equation}

The quadratic supertraces are given by (\ref{trace1})-(\ref{trace2}) and
\begin{align}
 &\langle \GG_{X'} \GG_{Y'} \rangle = \frac{n-2}{2}\delta_{X'Y'}\,, \qquad \langle \ZZ^2 \rangle = 4z_2^2 (4/d_n-1)\,,&\\
 &\langle \QQ^\alpha_{ij} \QQb^{kl}_\beta\rangle=-\delta^\alpha_\beta \Delta_{ij}^{kl}=-\langle \QQb^j_\beta \QQ^\alpha_i\rangle\,.&
\end{align}

\section{$SU(2,2|n)$ curvatures}\label{curvaturesApp}

Here we give the explicity expresions of the curvatures.

The field strength is defined by
\begin{equation}
\FF=\half \calF^{ab} \JJ_{ab}+\calF^a \JJ_a +\calG^a \KK_a +\calH \DD +\calF^I \TT_I +\calF \ZZ +\QQb_\alpha^i \calX_i^\alpha +\calXbar_\alpha^i \QQ_i^\alpha\,,
\end{equation}
\begin{align}
\calF^{ab}&=\mathcal{R}^{ab}-\psibar^i\se\Sigma^{ab}\se\psi_i\,\\
\calF^a&=Df^a+\frac{1}{s}g^a h+\frac{1}{2s}\psibar^i\se\gamma^a\se\psi_i\,,\\
\calG^a&=Dg^a+sf^ah-\half \psibar^i\se\tilde{\gamma}^a\se\psi_i\,,\\
\calH  &=H+sf^ag_a+\half \psibar^i\se\gf \se\psi_i\,,\\
\calF^I&= F^I-i\psibar^i\se(\lambda^I)_i^{\ j}\se\psi_j\,,\\
\calF &=F-\frac{i}{4 z}\psibar^i\se\se\psi_i\,,\\
\calX_i^\alpha &=D(\se \psi_i)^\alpha+\frac{s}{2}f^a (\gamma_a\se \psi_i)^\alpha +\half g^a(\tilde{\gamma}_a\se \psi_i)^\alpha+\half h (\gf \se\psi_i)^\alpha\,,\\
\calXbar^i_\alpha&=-(\psibar^i\se)_\alpha\overleftarrow{D}+\frac{s}{2} (\psibar^i\se\gamma_a)_\alpha f^a +\half (\psibar^i\se\tilde{\gamma}_a)_\alpha g^a+\half (\psibar^i\se\gf )_\alpha h\,,
\end{align}
where
\begin{align}
H&=dh\,,\\
\mathcal{R}^{ab}&=R^{ab}+s^2f^af^b-g^ag^b\,,\\
 R^{ab}&=d\omega^{ab}+\omega^a_{\ c}\omega^{cb}\,,\\
 F^I&=dA^I+\frac{1}{2} f_{JK}^{I}A^J A^K\,,\\
 F&=dA\,.
\end{align}
The covariant derivative on a Lorentz vector $V^a$ is 
\begin{equation}
 D V^a = dV^a+\omega^a_{\ b}V^b\,,
\end{equation}
and on a spinor
\begin{equation}
 D \psi^\alpha = d\psi^\alpha+\half \omega^{ab}(\Sigma_{ab}\psi)^\alpha-\frac{i}{2}A^I(\lambda_I\psi)_i^\alpha-i z(4/N-1)A\psi_i^\alpha\,. \label{covD}
\end{equation}

The covariant derivative $D$ is defined for the $SO(1,3)\times SU(n)\times U(1)$ connection. The left-acting exterior derivative satisfies $\Omega^m\overleftarrow{d}=(-1)^m d\Omega^m$ for an $m$-form, in the spinor representation yields the relation (\ref{covD}).

\end{appendices}

\bibliographystyle{ieeetr}
\bibliography{paper.bib}

\end{document}